# Large Alphabet Set Time-bin Encoded Measurement-Device-Independent Quantum Key Distribution

Gokul Agasthilingam, Harinee Natarajan, Varun Raghunathan[#]
Department of Electrical Communication Engineering
Indian Institute of Science
Bangalore 560012, India
[#]varunr@iisc.ac.in

Abstract: We report on the experimental demonstration of an expanded basis set (called here as alphabet set) time-bin encoded measurement-device-independent quantum key distribution (MDI-QKD). While MDI-QKD is known to prevent detector-side attacks, it inherently suffers from reduced secret key rate (SKR) due to coincidence measurements performed at the central measurement node. To address this limitation, we encode states across multiple time-bins thereby increasing possible coincidence events and mapping each successful alphabet exchange to multiple bits, thereby increasing the information capacity per alphabet transmitted. Using a standard MDI-QKD set-up with real fiber spools and single-photon avalanche photodetectors, we achieve SKRs of 401 (133.6) bps and 28 (10.7) bps for 8 (2) encoded states for distances of 2 and 50 km, respectively, resulting in 3- and 2.63-times improvement, respectively when compared to the conventional two-state encoding. Furthemore, the large alphabet set MDI-QKD results are compared with a similar encoding scheme implemented for coherent-one-way (COW) protocol. This comparison reveals a clear advantage of using a larger alphabet set for MDI-QKD, where increased Z-basis coincidence events yields increased SKR. These results provide important insights into the scalability of MDI-QKD key-rates without requiring additional hardware modifications, paving the way for next-generation, quantum key distribution networks.

## I. Introduction

The increase in data and associated cyber-security threats necessitates the use of robust cryptography solutions for securing data communication links. The security of existing classical cryptography solutions is based on computationally complex problems [1–3], which makes them increasing vulnerable to attacks with advancements in resource-efficient computational algorithms and their implementations in quantum computers [4–8]. Quantum cryptography solutions address this by achieving information-theoretic security for key exchange. Well-known quantum key distribution (QKD) protocols, such as BB84 and E91 [9,10], leverage fundamental principles of quantum mechanics for preparing and detecting quantum states with built-in capability for eavesdropper detection [11,12]. Despite the theoretical security guarantees, practical implementations of QKD often open unwanted side-channels for attacks owing to device imperfections such as non-ideal photon sources and detectors [13–18]. These potential side-channels can be exploited by adversaries to compromise the key exchange process without being detected. More specifically, detector side-channel attacks exploit imperfections in single-photon detectors, allowing an eavesdropper to intercept and manipulate signals without detection [19–22]. Detector-blinding attacks [22] can force the single-photon detector into classical mode while still mimicking legitimate detection events, thereby compromising on the security of the system. Measurement-device-independent QKD (MDI-QKD) addresses the detector-side vulnerabilities by removing the trust in the detection device thereby ensuring secure key exchange even with compromised detectors [23]. MDI-QKD implementation are also scalable to future quantum networks owing to their inherent star topology with multiple transmitter nodes communicating to a central measurement node.

Despite mitigating detector side-channel attacks, MDI-QKD faces practical limitations owing to its reliance on coincidence measurement for both key generation and eavesdropper detection. The probabilistic nature of the coincidence measurement at the central node combined with transmission losses across the quantum channel and the finite detector dead-time, limits the fraction of transmitted qubits contributing to useful key exchange [23–28]. For example, in a conventional qubit encoding scheme applied to MDI-QKD, only 12.5% of the transmitted states effectively contribute to useful key transmission, thereby limiting the overall system throughput. A promising solution to address this limitation is the use of high-dimensional (HD) quantum states, for information encoding [29–32]. HD time-bin encoding aims to improve QKD performance by partitioning the modulation time window into discrete time slots, with the position of the optical pulse in each time slot encoding a specific quantum state which gets mapped to multiple bits. HD encoding allows each encoded state to carry $log_2(d)$ bits of information, thereby increasing the information capacity per state transmitted. This approach mitigates the limitations imposed by detector dead-time and enhances robustness against channel losses. The HD time-bin encoding scheme is best suited for short to medium link lengths for which the detection rates are often limited by the detector dead-time [32].

HD encoding in the context of prepare-and-measure QKD protocols has been demonstrated with varying levels of hardware modifications [29–33]. The prepare-and-measure QKD protocols, however, are inherently prone to detector side-channel attacks, which remain unaddressed even with the HD implementation. In this context, HD-version of MDI-QKD protocol offers the combined benefits of preventing detector side-channel attacks and improving the secret-key throughput. A theoretical proposal for time-bin encoded HD version of MDI-QKD protocol has been reported previously [33]. However, practical limitations such as limited modulator extinction, detector efficiency, dead-time, after-pulsing, ark counts, and limited two-photon interference visibility have not been considered in this theoretical report.

In this work, we propose and experimentally demonstrate a practically viable large alphabet set MDI-QKD system for tim-bin encoding across multiple time slots. More specifically, we encode the quantum states across $d$ different Z-basis and two X-basis states, while relying on the multiple possible coincidence events for key exchange and error estimation. Using real fiber spools as the quantum channels and single photon avalanche detectors (SPAD), we experimentally achieve secret key rates (SKR) of 401 (133.6) bps and 28 (10.7) bps for 8 (2) state encoding for quantum channel lengths of 2 and 50 km, respectively. A clear improvement in SKR by as much as 3-times when compared to the two-state qubit implementation is achieved. For the experimental parameters considered here, we obtain an optimum alphabet set of 8 for which maximum SKR scaling is achieved, and beyond which SKR degrades due to increased bit and phase errors. We also compare the SKR scaling for the large alphabet versions of MDI- and coherent-one-way (COW) QKD protocol. The use of expanded alphabet set in MDI-QKD favourably improves the SKR scaling by: $\frac{2(d-1)}{d} log_2 d$ when compared to the standard $log_2 d$ scaling observed for COW-QKD. This improvement is ascribed to the increased coincidence possibilities across distinct time-bins which contibute to useful key exchange and hence the overall SKR.

## II. Theoretical Modeling

For the proposed large alphabet implementation of MDI-QKD protocol, Alice and Bob independently prepare time-bin encoded weak coherent states randomly selecting between the time-bin and phase bases, denoted as the Z- and X-basis, respectively. To prevent photon number splitting attacks [34], decoy states are also implemented by randomly choosing between three intensity states for X-basis encoding [35]. A step-by-step description of the proposed large alphabet set MDI-QKD protocol is given in Table 1. Z-basis states are realizedby partitioning the maximum available time window as determined by the clock cycle into $d$ different time slots allowing information to be encoded in the time of arrival of the photons. For illustration, a four-state encoding scheme is shown in Figure 1. The Z-basis states are represented by $|t_n\rangle$, which denotes optical pulse of duration $\Delta t$ in the $n^{th}$ time bin. The X-basis state, represented by $|f_n\rangle$, corresponds to the superposition of time-bin states with relative phase encoding between pulses. For the present implementation, we consider only two phase-basis states to simplify the implementation. The two X-basis states correspond to either zero or $\pi$ phase difference between pairs of alternating pulses across the d time-bin, as shown in Figure 1(b).

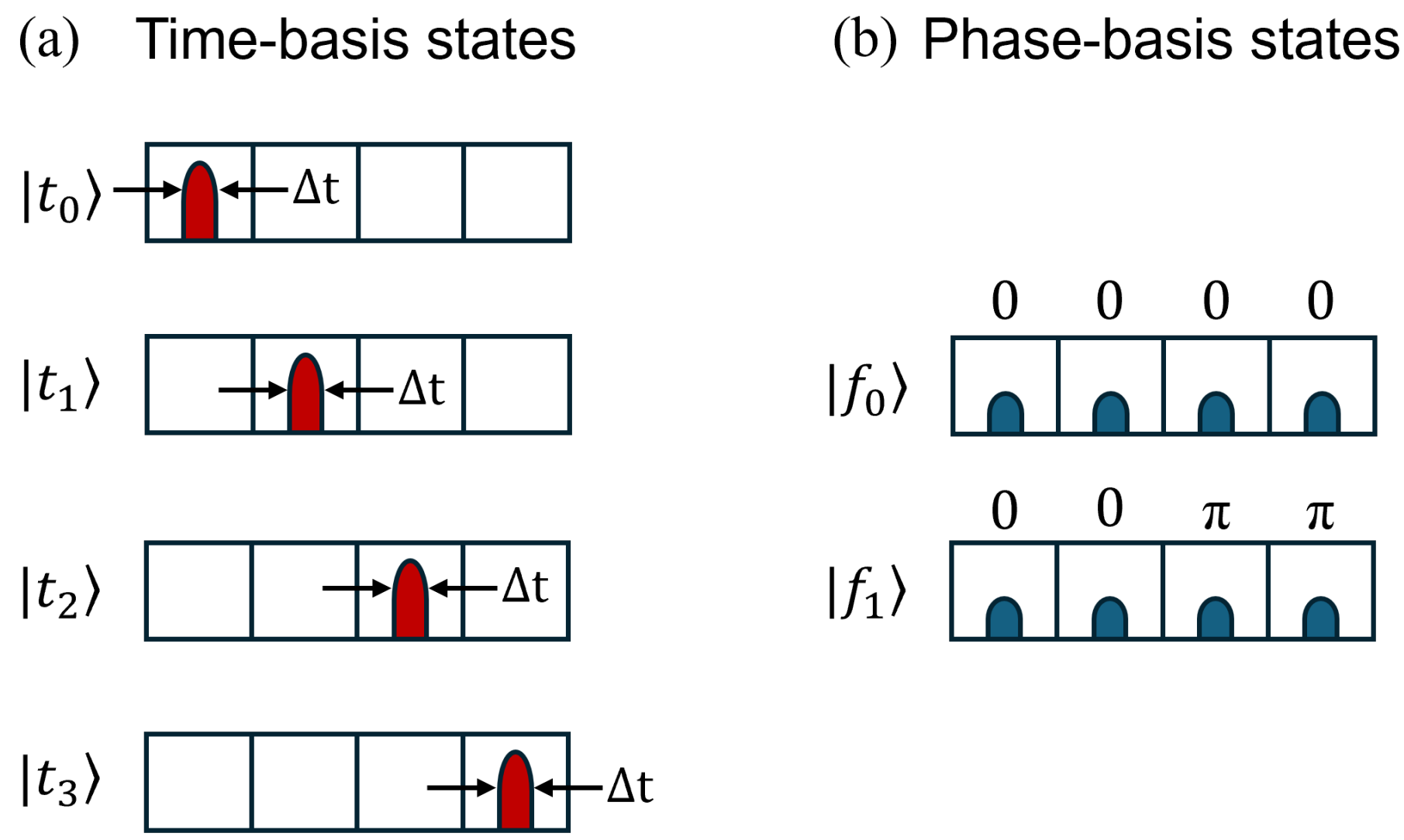

Figure 1: Illustration of time-bin encoded large alphabet set MDI-QKD with: (a) four encoded Z-basis states $\{|t_0\rangle, |t_1\rangle, |t_2\rangle, |t_3\rangle\}$, and (b) two encoded X-basis states $\{|f_0\rangle, |f_1\rangle\}$.

This ensures mutually unbiased basis outcomes with respect to the Z-basis states, such that $\langle t_n|f_n\rangle = 1/d$. A security proof for prepare-and-measure QKD implementation employing a reduced number of monitoring (X-basis) phase states has been reported previously [36]. It is noted that reducing the number of phase basis states does not compromise security in the context of prepare-and-measure scenario, although it can limit the achievable communication distance since the X-basis phase-error rate ($e_x$) increases relative to the Z-basis bit-error rate ($e_z$). An unconditional security proof for MDI-QKD using reduced X-basis states has not been established. Extending the analysis to a full d-state X-basis within the MDI framework also remains an open direction for future work**.**

Alice and Bob select with equal probability the basis for encoding (50:50 between Z- and X-basis) and the actual states to be encoded (*d* and 2 states for Z- and X-basis, respectively). The time-bin encoded states are subsequently transmitted to a central, untrusted intermediary relay named, Charlie, for coincidence measurements. As shown in Figure 2, the detection events consist of coincidence detection between distinct time-bin pairs, represented as: $\{(|d_{1i}\rangle, |d_{2j}\rangle)$ or $(|d_{1j}\rangle, |d_{2i}\rangle)\}$, with detectors D1 and D2 clicking in the $i^{th}$ and $j^{th}$ time-bin, respectively or vice versa. The coincidence events across the two detectors are termed here as *level-d coincidences*, which are operational signatures of anti-symmetric two-photon interference across different time bins. Charlie, the untrusted relay, publicly announces the level-d coincidence events with information about which detector clicked and the corresponding time slots. For a d-dimensional system, there exist *d(d-1)* possible level-d coincidence events. This results in increased key transmission in comparison to standard two-

*Table 1: Description of the proposed large alphabet set MDI-QKD protocol*

1. **State preparation:** For each block of transmission, Alice and Bob generate weak coherent pulses, randomly choosing between Z- and X-basis states: $|t_i\rangle$ and $|f_j\rangle$, with i = 0 to $d$-1, and j = 0 to 1 state, respectively. For Z-basis, only signal states are encoded in the respective time bins are generated. For X-basis, decoy states are encoded with mean photon numbers $q_a \in \{\mu_a, \nu_a, \omega_a\}$ and $q_b \in \{\mu_b, \nu_b, \omega_b\}$ with probabilities: $P_{q_a} \in \{P\mu_a, P\nu_a, P\omega_a\}$ and $P_{q_b} \in \{P\mu_b, P\nu_b, P\omega_b\}$ at Alice and Bob, respectively. μ, ν, and ω represent the signal, decoy, and vacuum states, respectively. These states are transmitted to the central measurement node, named Charlie.
2. **Measurement:** The states sent by Alice and Bob are interfered at Charlie's end with the announcement of level-d coincidences indicating detection clicks for detectors D1and D2, and the time slot for which the detection occurred, denoted as: $\{d_{11}, d_{12}\ldots d_{1d}\}$ and $\{d_{21}, d_{22}\ldots d_{2d}\}$, respectively. The level-*d* coincidences are the operational signatures of anti-symmetric two-photon interference across distinct time bins.
3. **Basis sifting:** Alice and Bob announce the basis used to encode the pulses for which Charlie announces successful level-*d* coincidence events. As in the case of conventional MDI-QKD protocol, Bob flips his state to match with that of Alice when both transmit Z-basis or X-basis states. The states are correspondingly mapped to multiple bits with d-time slots mapped to $log_2 d$ bits. For X-basis transmissions, only opposite phase patterns contribute to coincidence signatures with same-phase pairs contributing to error and are used for parameter estimation.
4. **Parameter estimation:** Alice and Bob share a fraction of their sifted states for estimating the error rates for the Z- and X-basis. Here, the Z basis is used to generate key bits and X basis is used for monitoring the presence of an eavesdropper. Using the decoy states for X-basis, the single photon gain ($Q_{11}^z$) and error rates ($e_{11}^x$) are calculated and used for SKR estimation.
5. **Key distillation:** Error correction, verification, and privacy amplification steps are performed to ensure a common secret key is distributed between Alice and Bob.

state qubit implementation of MDI-QKD. Same-slot detection events (i = j) are discarded, as such outcomes reveal the encoded time slot to Charlie. For the X-basis, which is the monitoring basis, the pulses are phase-tagged within each time bin, as shown in Figure 3. The detection

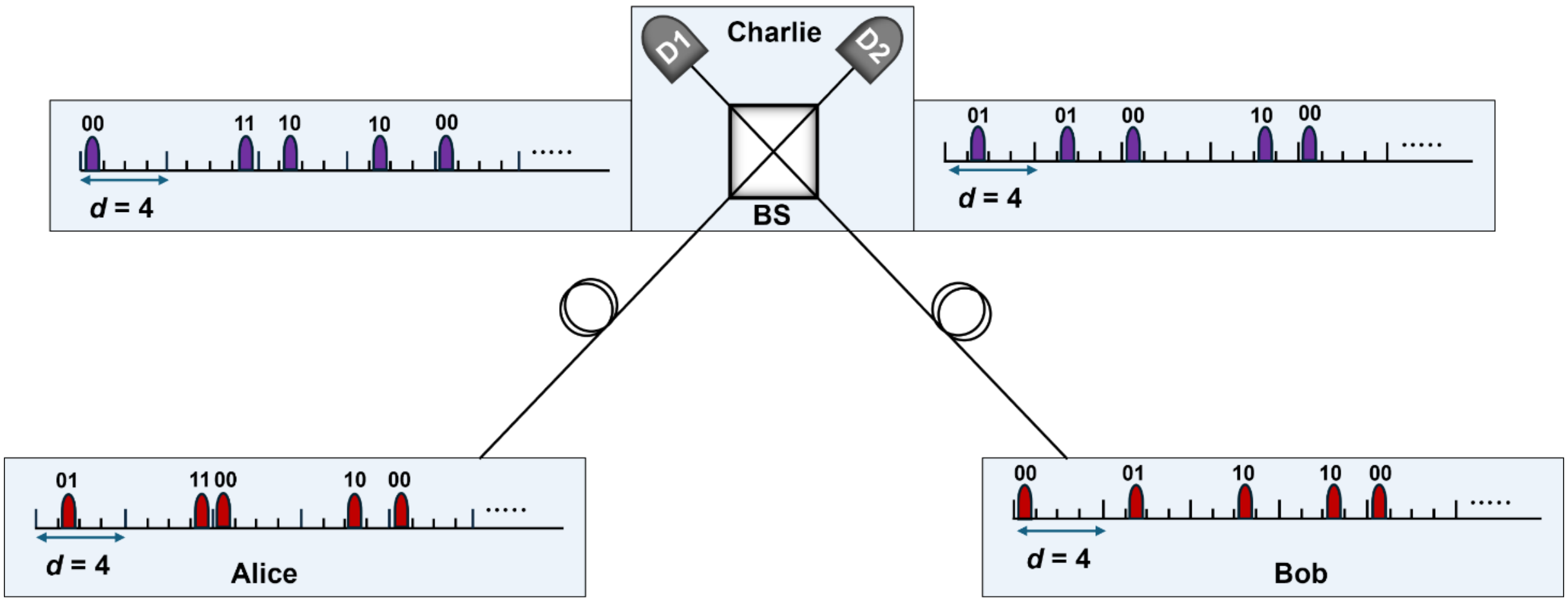

Figure 2: Illustration of MDI-QKD transmission for four state encoding at Alice and Bob, with coincidence measurements performed at Charlie. The level-d coincidence measurement outcomes correspond to detection events of the form: $\{|d_{1i}\rangle|d_{2j}\rangle\}$, with clicks at i-th and j-th time-bins for detectors $D_1$ and $D_2$, respectively.

pairs are analyzed to distinguish whether Alice and Bob transmitted different states among $\{|f_0\rangle, |f_1\rangle\}$. Considering four-state time-bin encoding as an example, the different level-*d* coincidence events for the transmission of X-basis states are: (i) detection corresponding to $|d_{11}\rangle$ or $|d_{12}\rangle$ occurs, then look for coincidence detections corresponding to $|d_{23}\rangle$ or $|d_{24}\rangle$, or (ii) detection corresponding to $|d_{13}\rangle$ or $|d_{14}\rangle$ occurs then, look for coincidence detections corresponding to $|d_{21}\rangle$, or $|d_{22}\rangle$. Once the successful level-d coincidence detection events are

announced, both Alice and Bob reveal the basis used for encoding. Similar to the conventional MDI-QKD, Bob flips his state to match with that of Alice when their transmission bases are aligned. Any deviation from this, especially when same states are transmitted resulting in level-d coincidence contribute to the error in transmission of both Z and X basis states. The decoding strategy proposed here is inherently scalable to larger alphabet set with increased key throughput.

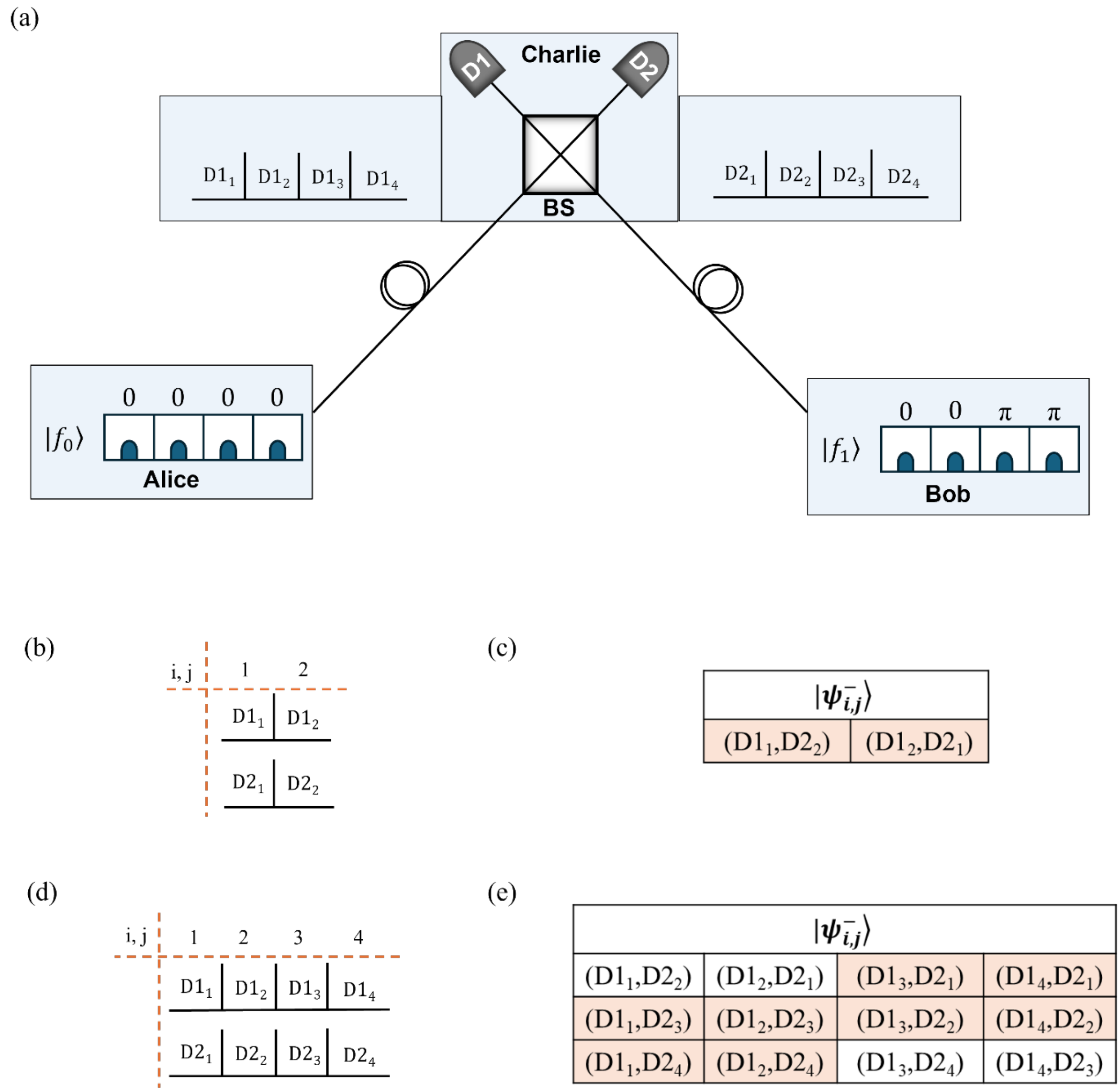

Figure 3 (a) Generation and detection of encoded states in X-basis with four states encoded. (b) Detection times slots for two-state encoding. (c) Possible detection pairs for extracting level-d coincidence pairs for two-state encoding. (d) Detection time slots for four-state encoding. (e) Possible detection pairs to extract level-d coincidence pairs for four-state encoding. The useful events for extracting level-d coincidence are shown as orange-shaded cells in (c) and (e).

The Z basis states transmitted by Alice and Bob with optical pulses in the $i^{th}$ *and* $j^{th}$ time-bins with $i \neq j$ result in projections onto two-dimensional symmetric and antisymmetric states of the form: $|\psi^{\pm}_{i,j}\rangle = \frac{1}{\sqrt{2}}(|\, i\rangle_A\, |\, j\rangle_B \pm |\, j\rangle_A\, |\, i\rangle_B)$ [23]. The antisymmetric state detection events result in level-d coincidences, which are considered useful for key exchange. The symmetric state detection resulting in same detector clicking in different time slots are discarded in the present implementation due to detector deadtime limitations [35]. For the X basis measurements, we use four-dimensional encoding as an illustrative example here. The phase encoding produces superpositions of time-bin states with either same or alternating 0-π phase pattern between consecutive pulse pairs, defined here as: $|f_0\rangle = \frac{|0\rangle + |1\rangle + |2\rangle + |3\rangle}{2}$ and $|f_1\rangle = \frac{|0\rangle + |1\rangle - |2\rangle - |3\rangle}{2}$. For two different X-basis states transmitted by Alice and Bob of the form $|f_0\rangle|f_1\rangle$ and $|f_1\rangle|f_0\rangle$, projection onto two-dimensional Bell state can be written of the form: $\frac{1}{2\sqrt{2}}\left[|\Phi_{01}{}^{+}\rangle + |\psi_{01}{}^{+}\rangle + |\psi_{02}{}^{-}\rangle + |\psi_{12}{}^{-}\rangle + |\psi_{03}{}^{-}\rangle + |\psi_{13}{}^{-}\rangle + |\psi_{23}{}^{+}\rangle + |\Phi_{23}{}^{+}\rangle\right]$. Here, the $|\Phi_{ij}{}^{\pm}\rangle$ *and* $|\psi_{ij}{}^{\pm}\rangle$ denote the four Bell states in two-dimensional basis. Note the presence of both antisymmetric and symmetric states in the expansion. The antisymmetric states result in level-d coincidence between time-slots 0 or 1 and 2 or 3 with a probability of ½. The symmetric states result in same detector clicking at different time slots and are discarded in the present implementation. For the same X-basis states transmitted by both Alice and Bob of the form $|f_0\rangle|f_0\rangle$ and $|f_1\rangle|f_1\rangle$, the projection onto two-dimensional Bell state can be written of the form: $\frac{1}{2\sqrt{2}}\left[|\Phi_{01}{}^{+}\rangle + |\psi_{01}{}^{+}\rangle + |\psi_{02}{}^{+}\rangle + |\psi_{12}{}^{+}\rangle + |\psi_{03}{}^{+}\rangle + |\psi_{13}{}^{+}\rangle + |\psi_{23}{}^{+}\rangle + |\Phi_{23}{}^{+}\rangle\right]$ and $\frac{1}{2\sqrt{2}}\left[|\Phi_{01}{}^{+}\rangle + |\psi_{01}{}^{+}\rangle - |\psi_{02}{}^{+}\rangle - |\psi_{12}{}^{+}\rangle - |\psi_{03}{}^{+}\rangle - |\psi_{13}{}^{+}\rangle + |\psi_{23}{}^{+}\rangle + |\Phi_{23}{}^{+}\rangle\right]$, respectively. All the above terms represent symmetric states, which are also discarded. The observations above can be generalized to larger alphabet encoding resulting in specific level-d coincidence outcomes across time-bins based on the relative phase pattern across the interfering X-basis pulses, as discussed in the supplementary material, section S1.2.

It is noted that the measurements performed at Charlie's end are inherently restricted by the limitations imposed by linear optics. As established in Vaidman et. al. [37] and later quantified in Casamiglia et. al. [38], a Bell-state analyzer employing only passive linear-optical elements and two detectors can unambiguously distinguish at most two of the four Bell states, giving a maximum theoretical success probability of 50%. Consequently, our receiver comprising of a 2 × 2 beam combiner and two single-photon detectors implement partial projections onto the symmetric and antisymmetric two-photon states and cannot perform the complete $d \otimes d$ Bell-state measurements. These measurements are same as the two-dimensional MDI-QKD with projections onto two-dimensional Bell states, as shown above. Furthermore, as shown by Lo et. al. [23], successful projections onto one of the Bell states provides the necessary correlations for key exchange in a two-state MDI-QKD scheme, while all other detection outcomes are treated as inconclusive. This is routinely utilized in practical implementations of two-dimensional MDI-QKD [23-27,35], confirming that full Bell-state discrimination is not required for MDI-QKD operation.

In this work, we follow the general MDI-QKD security framework established in Ref. [23, 39], where coincidence observables are used to bound the single-photon gain and phase-error rates through decoy-state analysis. In our experiment, each level-d coincidence corresponds to antisymmetric projections between two distinct time-bin modes, which can be considered operationally identical to the Bell-state measurement in conventional two-state MDI-QKD and thus provides the required entanglement correlation for MDI security [23]. Hence, we conclude that the large alphabet set MDI-QKD protocol studied here can be as secure as the two-dimensional implementation [23,24,35] providing the required immunity against detector side-channel attacks.

Although the measurements described above are developed for the single-photon states generated at Alice and Bob, it can be extended naturally to practical implementations using phase-randomized weak coherent pulses (WCPs) [23-27,35]. Each transmitted pulse can be regarded as a superposition of number states, and the interference between the single-photon components of Alice's and Bob's WCPs at the beam combiner results in same Bell-state projections as in the single-photon scenario. The experimentally observed coincidences with WCPs are used to bound single photon gain and error rates through decoy-state method. This approach is consistent with the standard MDI-QKD security framework developed for the use of WCPs [23, 35, 37]. The decoy-state analysis and final secret key generation utilized here are hence expected to be compliant with existing security proofs for conventional, two-dimensional MDI-QKD [23, 39].

Building upon the mathematical formalism proposed in ref. [39], a theoretical model for the expanded alphabet set MDI-QKD is developed considering $d$ Z-basis states and two X-basis states. The model considers various system parameters which influence the achievable SKR such as, detector efficiency, dead time, dark counts, after-pulsing and background noise. Further discussion on the theoretical model is provided in the supplementary material, section S1.1. The overall SKR derived for the proposed MDI-QKD scheme is obtained as:

$$S = Q_{11}^{z}\left(log_2 d - h_d(e_{11}^{x})\right) - leak_{EC}, \tag{2}$$

where $h_d$ represents the $d$-dimensional Shannon entropy which is given as [40]:

$$h_d(Q) = -Q log_2 d\left(\frac{Q}{d-1}\right) - (1-Q) log_2 d(1-Q), \tag{3}$$

$Q_{11}^{z}$ is the single-photon coincidence gain, i.e. the probability conditioned on single-photon emissions by Alice and Bob that Charlie reports any level-d coincidences in the announced frame. The phase-error $e_{11}^{x}$ is inferred from X-basis coincidence statistics using the decoy-state bounds. For each intensity pair ($q_a$, $q_b$) chosen from $q_a \in \{ \mu_a, \nu_a, \omega_a \}$ and $q_b \in \{ \mu_b, \nu_b, \omega_b \}$ with probabilities, $P_{q_a} \in \{ P\mu_a, P\nu_a, P\omega_a \}$ and $P_{q_b} \in \{ P\mu_b, P\nu_b, P\omega_b \}$, and basis $B \in \{z,x\}$, we estimate the coincidence gain $Q_{q_a q_b}^{B} = \sum_{i \neq j} C_{ij}^{B} / N_B$. and error $e_{q_a q_b}^{B}$ from the observed level-d coincidences $C_{ij}^{B}$ over $N_B$ transmitted frames. This definition includes successful detection events regardless of whether Alice and Bob send same or different states. The error rate, $e_{\mu_A \mu_B}^{B}$, is the fraction of detection event resulting in incorrect bit assignment. For the case where $i \neq j$, coincidence events corresponding to the state $|\psi_{i,j}^{-}\rangle$ can still occur owing to spurious detection events. As these events originate for detections at different time bins than those used for state preparation, they lead to erroneous detections and thus contribute to the overall QBER. Errors can also occur when Alice and Bob send quantum states in the $i^{th}$ and $j^{th}$ time bins with $i = j$, resulting in level-d coincidences. The SKR and overall system performance for the proposed MDI-QKD scheme depends on the level-d coincidence events. The ideal key-rate scales with

(a)

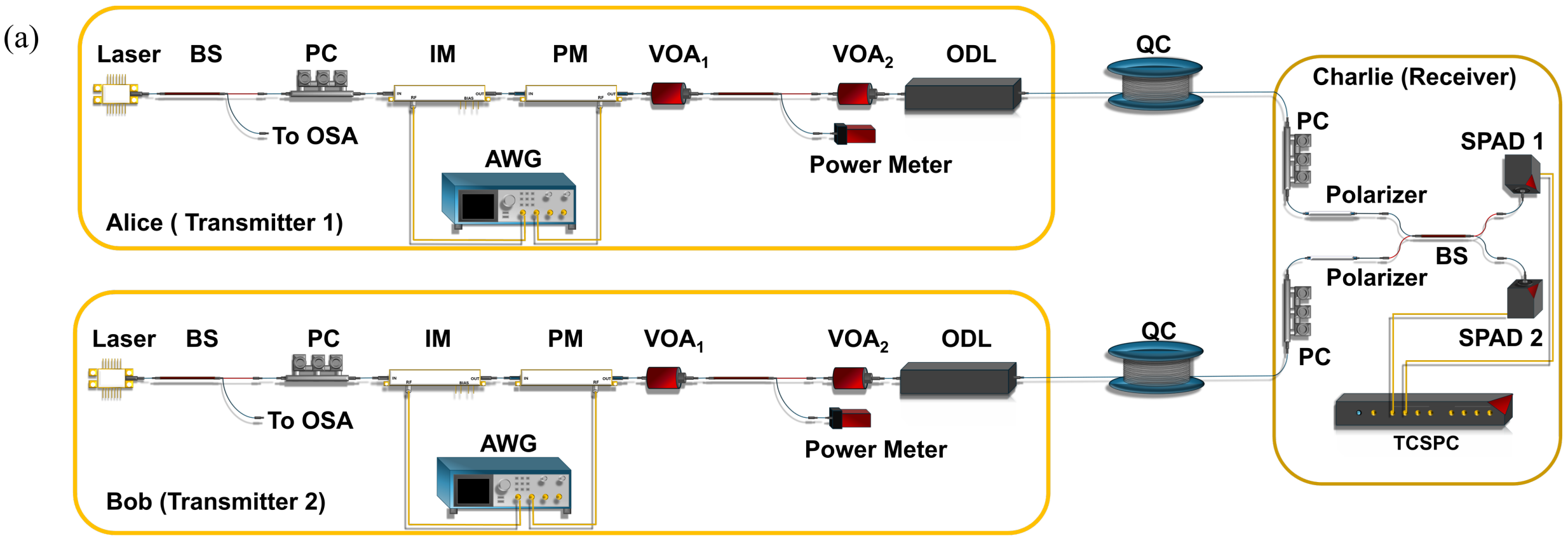

(b)

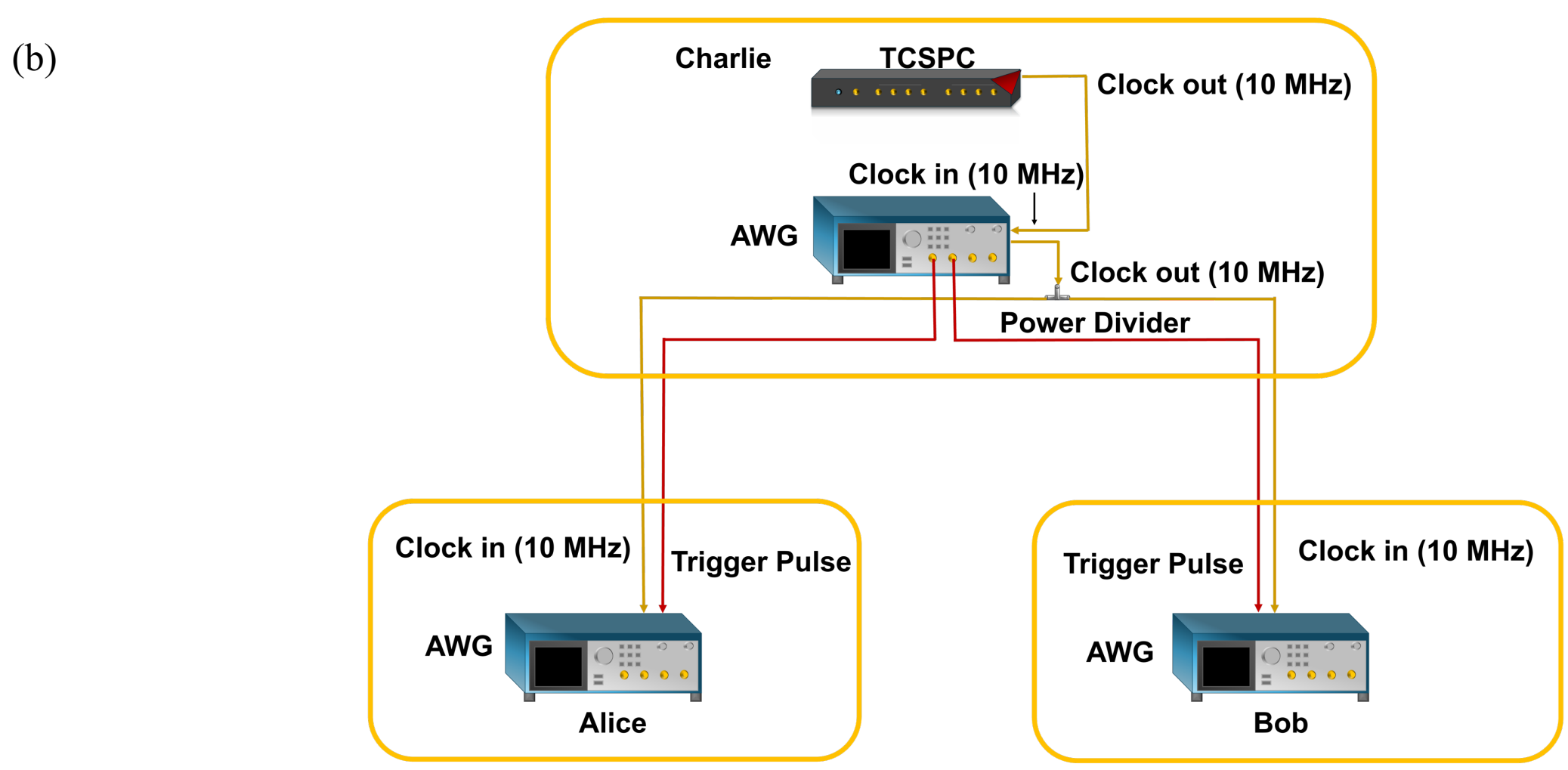

Figure 4: (a) Schematic of the MDI-QKD experimental setup. Continuous wave lasers at Alice and Bob are modulated using intensity modulator (IM) and phase modulator (PM). A 90:10 beam splitter directs 10% of the optical power to an optical spectral analyzer (OSA) for wavelength monitoring. The modulated pulses are then sent to variable optical attenuators (VOAs) to set the mean photon number. An optical delay line (ODL) ensures temporal alignment of pulses as they arrive at Charlie. The pulses received from both Alice and Bob are combined at Charlie's end using a 2x2 beams combiner after matching their polarization states using motorized polarization controller (PC) and inline polarizers. The outputs of the beam splitters are connected to single-photon avalanche detectors (SPADs) for coincidence measurement. (b) Experimental setup for synchronization of the three nodes (Alice, Bob and Charlie). Charlie's TCSPC clock is synchronized to an AWG, which serves as the master waveform generation to synchronize the clock at Alice and Bob end. A power divider at Charlie's AWG output is used to distribute the clock signals, which are fed into clock inputs of Alice and Bob AWGs, thereby implementing three-node synchronization. Trigger pulses are also distributed from Charlie to Alice and Bob for commencing the quantum key transmission

encoding dimension $d$ as: $\frac{2(d-1)}{d} log_2\, d$. As $d$ increases, the fraction of redundant detection events, i.e., when coincidences are recorded for the same time-bin decreases. This scaling of useful coincidence events with $d$ underpins the improvement in secret-key throughput with increased alphabet set for MDI-QKD.

## III. Experimental Implementation

### A. Transmitter Node:

A schematic of the transmitter nodes used for the MDI-QKD experiments is shown in Figure 4(a). The two transmitter nodes, labelled Alice and Bob, generate the time-bin encoded states with near indistinguishable outputs in terms of center wavelength, polarization state, and time of arrival at the central measurement node. Each transmitter uses a narrow-linewidth, fiber coupled diode laser (Eblana Photonics, DFB laser), operating at a center wavelength of ~1550 nm. A 90:10 optical beam splitter is used to pick-off 10% of the optical power to measure the laser spectrum using an optical spectrum analyzer (OSA, Yokogawa AQ6370D) for wavelength monitoring. The remaining 90% is sent to an intensity modulator (IM, iXBlue MXER-LN-10) for time-bin encoding and subsequently to a phase modulator (PM, iXBlue, MPZ-LN-10) for relative phase encoding between the time-bin encoded pulses. Four separate output sequences from a home-built quantum random number generator [41] are used for selecting the bases, encoding states, relative phase values and decoy states. The corresponding electrical waveforms for modulating the optical pulses and relative phase setting are obtained from arbitrary waveform generators (AWG, Tektronix AWG5204) and amplified to the required voltage levels before applying to the IM and PM. The states are encoded at a clock rate of 10 MHz with 1 ns separation between time-bins and ~600 ps pulse-width. The mean photon number per pulse is precisely controlled using two electronic variable optical attenuators (EVOA, Thorlabs VOA50-APC). At the output of the first EVOA, a 90:10 beam splitter is used to direct 10% of the optical power for power monitoring, while the remaining 90% undergoes additional attenuation using the second EVOA to achieve the required mean photon level. This configuration ensures accurate preparation of the signal and decoy states,with mean photon numbers set to $u$ = 0.2 (signal state), $v$ = 0.12 (decoy state), and $w$ = 0 (vacuum state) for X-basis states. The mean photon number for the signal and decoy states are chosen based on an optimization model used to maximize the overall SKR [39]. A motorized optical delay line (ODL, Agiltron, MDTD-01C111323) is used to precisely compensate for any temporal misalignments between the generated optical pulses before propagating through standard single mode fiber spools toward the central measurement node. The overall stability of the transmitter nodes, in particular, the mean photon number, polarization state, and center wavelength setting over six hours duration is shown in Section S3 of the supplementary material.

### B. Measurement Node:

A schematic of the central measurement node (relay), labelled Charlie, used for performing coincidence measurements on the transmitted states, is shown in Figure 4(a). The polarization states of the incoming optical pulses are matched using motorized polarization controllers (PCs, Thorlabs MPC220) and inline fiber polarizers (Thorlabs ILP1550PM-APC). The two incoming states are combined using a 2×2 fiber-based beam combiner (Thorlabs PN1550R5A2) to enable two-photon interference at the central measurement node. The two outputs of the beam combiner are connected to a single-photon avalanche detector (SPAD, IDQuantique IDQube-NIR), operated in gated mode with a typical detection efficiency of 25%, deadtime of 10 µs, and a maximum dark count rate of 2 kHz. Gate pulses with a temporal pulse width of *(d+3)* ns are generated using a time-correlated single-photon counting module (TCSPC, IDQuantique ID900). The SPAD outputs are fed to the TCSPC system for time-tagging and recording photon-detection events. For the present lab-level demonstration, standard electrical cables (BNC or SMA) are used for classical communication, for system synchronization, clock, and trigger signal distribution. These channels can readily be replaced by optical fibers for field-level implementations. Owing to the inherent signal arrival time difference between the classical and quantum signals, a post-processing step is performed on the recorded timestamps to compensate for temporal offsets before performing parameter and key estimation.

**C. Three-Node Synchronization:**

Accurate mapping of the level-*d* coincidence measurements for useful key exchange requires tight synchronization between Alice, Bob, and Charlie. This requires the waveforms generated at Alice and Bob to be synchronized and triggered simultaneously, as shown in Figure 4(b). A master waveform generator is used at Charlie's node, with its internal clock locked to the 10 MHz reference output of the TCSPC. The master waveform generator distributes a 10 MHz reference clock to Alice and Bob for clock synchronization. Additionally, two output channels of the master waveform generator are used to generate trigger signals for the waveform generators at Alice and Bob to commence the key distribution experiment. The timing skew between the trigger signals are carefully adjusted to ensure simultaneous generation of electrical waveforms at both Alice and Bob.

**D. MDI-QKD Experiments:**

The performance of the MDI-QKD system is characterized using single-mode optical-fiber spools as the quantum channel with three different lengths and four different equivalent loss values: two 1 km fibers (with 1.62 dB total loss), two 25 km fibers (with 10 dB total loss), two 50 km fibers (with 22 dB total loss) and another set of two 50 km fibers (with 30 dB total loss). To characterize the extent of indistinguishability of the incoming photons at the central measurement node, we performed Hong-Ou-Mandel (HOM) interference for the weak coherent pulse inputs by measuring the coincidence counts as a function of varying optical delay, as shown in Figure 5(a). HOM visibilities in the ranges of 45–49% are obtained for the weak coherent states generated. Furthermore, we perform the HOM interference experiment as a function of wavelength difference between Alice and Bob's laser sources, as shown in Figure 5(b). A sharp HOM dip is observed when the wavelengths of the two laser sources are aligned, with a full width at half maximum of 6.33 pm. The two photon interference experiments indicate the need for precise time-delay and wavelength alignment for the X-basis detection in the MDI-QKD experiments requiring a wavelength overlap to within 1.8 pm. The stability of the HOM interference as measured using the visibility metric acquired over six hours duration is shown in section S3 of the supplementary material.

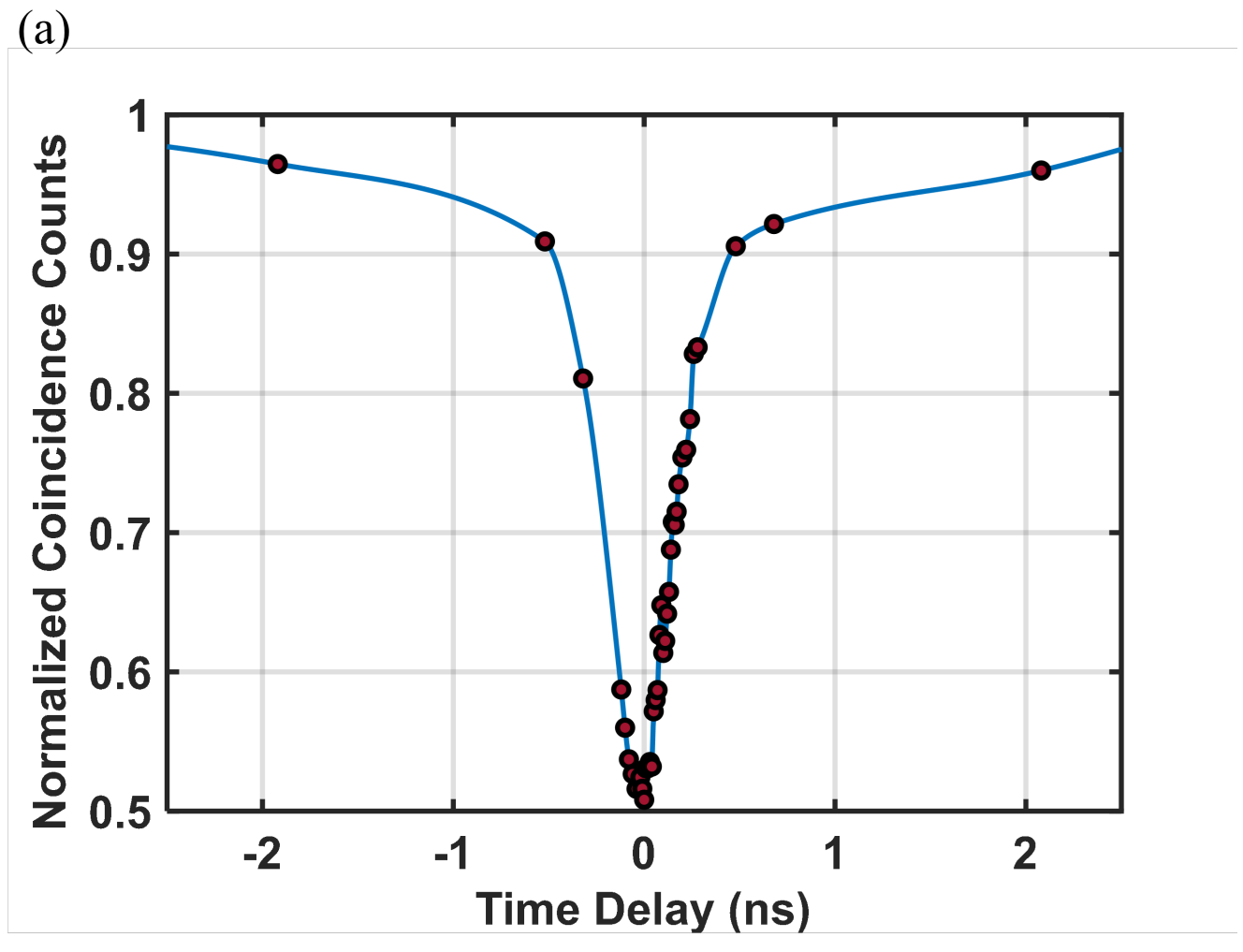

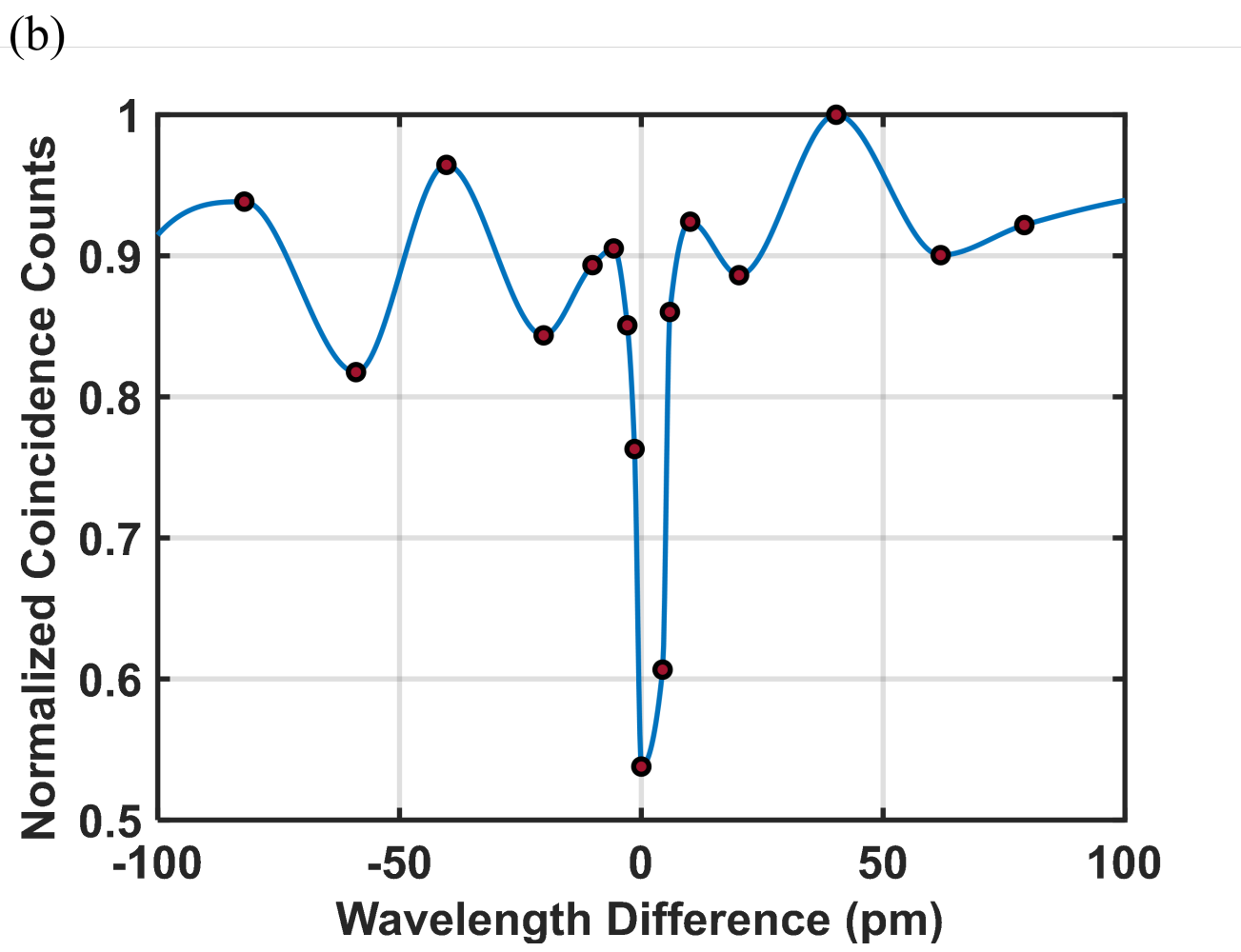

Figure 5: (a) HOM interference for the MDI-QKD experiment shown as the normalized coincidence versus the temporal delay between the two optical pulses arriving from the two transmitter nodes. (b) HOM Interference shown as the normalized coincidence as a function of wavelength difference between laser sources at Alice and Bob. The green data points represent the experimental and blue curve represents fit to the data.

The SKRs obtained after subtracting the eavesdropper's information, as a function of link length for the MDI-QKD experiments with $d = 2$, 4, and 8 state encoding, are shown in Figure 6(a). For shorter distances, a clear increase in SKR with $d$ is observed, which can be attributed to the scaling due to level-$d$ coincidences contributing to increased useful key generation. The SKR numbers are increased by 2.63 times and 3-times are observed for d = 8 and 2, respectively, when compared to the two-state implementation for a total fiber length of 50 km. Beyond this distance, the key-rates for d > 2 implementations decrease compared to the qubit transmission, with negative SKR obtained for 100 km total length. This deterioration in SKR is attributed to an increased bit- and phase-error rate due to significant contribution from imperfect modulator extinction, detector dark counts and misalignment errors when compared to the useful detection events. Both experimental and simulation results point to an optimum number of encoding states for which maximum SKR is achieved. The SKR as a function of encoding dimension d for 50 km total length is shown in Figure 6(b). For the present MDI-QKD implementation, an optimum encoding state number of 8 is obtained for both 2 and 50 km channel lengths between Alice to Bob. A summary of the experimental results for the large alphabet set MDI-QKD including the SKR obtained after subtracting the eavesdropper's information, the corresponding quantum bit error, and phase error rates, and the respective gains and error rates used for SKR estimation is included in section S4 of the supplementary material.

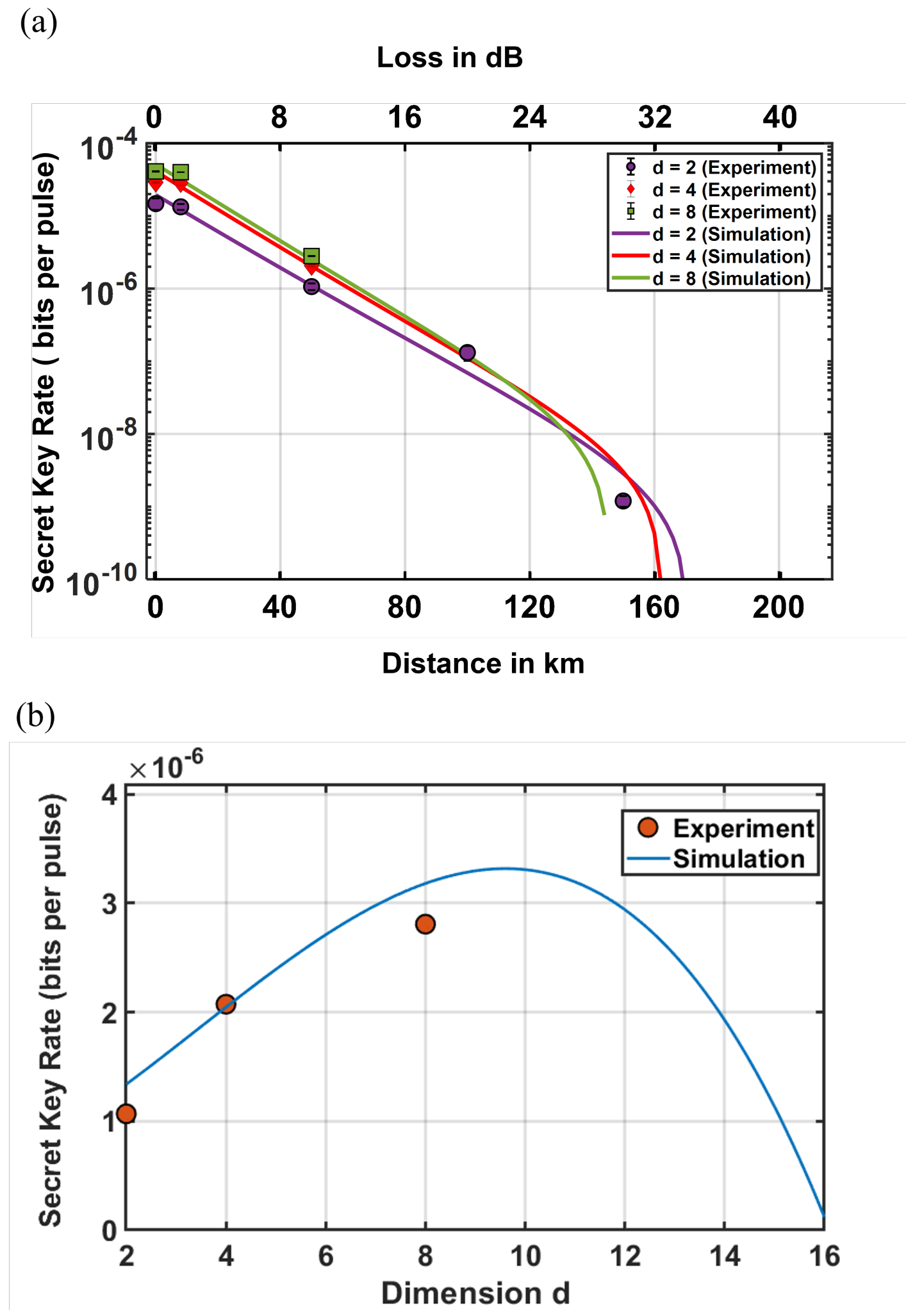

Figure 6: (a) SKRs obtained in units of bits per pulse as a function of total link length from Alice to Bob for MDI-QKD experiments with d = 2 (purple data points), d = 4 (red data points), and d = 8 (green data points) encoded states. (b) SKRs obtained for a fixed 50 km total channel length as a function of the number of the encoded states. The solid curves represent the theoretical fits to the experimental data.

The present implementation uses a fixed clock rate of 10 MHz for demonstrating SKR scaling with increased alphabet set for time-bin encoding. To understand the role of increasing the clock rate for SKR scaling for both two-state and expanded alphabet set encoding, we model the normalized SKR as a function of normalized clock rate under both ideal and realistic (non-ideal) conditions. These results are included in section S5 of the supplementary material. The realistic QKD model considers transmission losses, limited detector efficiency, dark counts, after-pulsing probability, limited modulator extinction, and reduced HOM visibility; all of which contribute to deterioration in SKR due to increased errors. For a fixed clock rate, we observe an increase in SKR with increasing number of encoded states, and similarly for a fixed alphabet set we observe an increase in SKR with increasing clock-rate. The ideal model also points to a 50% increase in SKR for the for-state encoding when operated at half the maximum clock rate. This SKR scaling however degrades when considering in real QKD implementation.

**E. Comparison Between MDI-QKD and COW-QKD Implementations:**

Next, we compare the SKR scaling as a function of the number of encoded states for MDI-QKD and COW-QKD protocols to assess the usefulness of incorporating an enlarged alphabet set for time-bin encoding. Previously, we have implemented COW-QKD using multiple time-bin encoded states [42] with d different Z-basis states and (d-1) X-basis states achieving 1.694-times improvement in SKR when compared to the two-state COW-QKD with a maximum SKR of 210 kbps obtained for d = 16 across 25 km link length [40]. The present comparison uses MDI- and COW-QKD results obtained for nearly identical channel lengths of 50 km (equivalent to 10 dB losses) and 58.79 km (equivalent to 11.75 dB loss) [43] between Alice and Bob, respectively. The conclusions drawn here would not change if identical fiber lengths were used for MDI- and COW-QKD implementations. Figure 7(a) shows the theoretical plots of normalized SKR as a function of the encoding dimension with the normalization performed with respect to the two-state encoding scheme. The theoretical models consider both the ideal scaling without any device imperfections (magenta and green curves) and the realistic non-ideal case (blue and red curves). The models used for extracting SKR for the two protocols are described in sections S1 and S2 of the supplementary material. The corresponding experimental results obtained for the two protocols are also shown in Figure 7(a) - blue and orange data points. The normalized SKR as a function of number of encoded states is consistently higher for MDI-QKD compared to COW-QKD. A summary of the normalized SKR values obtained from the ideal theoretical, non-ideal simulation model, and experimental results is also tabulated in Figure 7(b).

For the ideal model, a monotonic increase in the normalized SKR is observed with increasing $d$, with the MDI-QKD key-rate scaling more rapidly when compared to that for the COW-QKD case. For eight-state encoding, the normalized SKR values are 3 and 5.25 for COW- and MDI-QKD, respectively. This scaling can be attributed to the $\frac{2(d-1)}{d} log_2(d)$ improvement in key-rate for MDI-QKD owing to the increased level-d coincidence events, while conventional prepare-and-measure scheme scale as $log_2(d)$ [29–32]. The non-ideal simulation model and experimental results show deviations from ideal plots due to practical limitations arising from device imperfections and transmission losses. This result in the SKR reaching a maximum value for an optimal number of encoding states beyond which the improvement deteriorates. Our experiments achieve a SKR improvement of 2.63 for eight-state encoded MDI-QKD when compared to the two-state implementation. The reduction in normalized SKR for both MDI

(a)

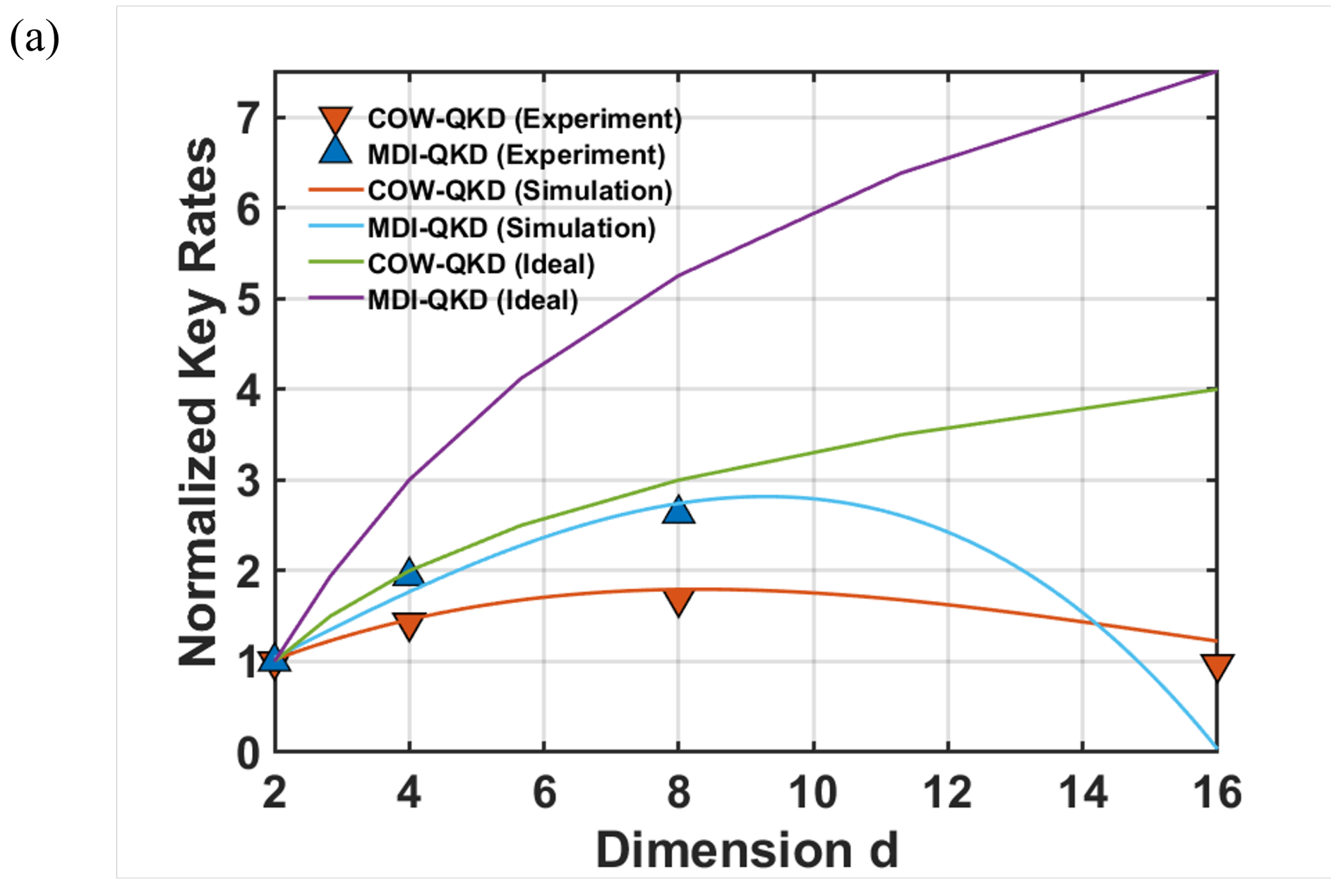

(b)

| COW-QKD : $log_2 d$ | | | | MDI-QKD : $\frac{2(d-1)}{d} log_2 d$ | | |
|---|---|---|---|---|---|---|
| d | Ideal | Simulation | Experiment | Ideal | Simulation | Experiment |
| 2 | 1 | 1 | 1 | 1 | 1 | 1 |
| 4 | 2 | 1.508 | 1.426 | 3 | 1.865 | 1.942 |
| 8 | 3 | 1.723 | 1.694 | 5.25 | 2.596 | 2.63 |

Figure 7: (a) Comparison of the normalized SKR as a function of d considering ideal case for MDI-QKD (purple curve), and COW-QKD (green curve), simulated non-ideal case for MDI-QKD (blue curve), and COW-QKD (red curve) and experimental results for MDI-QKD in (blue data points) and COW-QKD (red data points). (b) Table summarizing the normalized SKR for the ideal, non-ideal simulations, and experiments.

and COW experiments with further increase in d is attributed to increased error-rates resulting in larger fraction of the raw transmitted bits being utilized for error correction. Both theoretical and experimental results clearly point to MDI-QKD benefitting from the extended alphabet set used for encoding due to the increased level-$d$ coincidence events at the central measurement node improving the overall SKR scaling.

## IV. Conclusions:

In this paper, we experimentally demonstrate large alphabet set MDI-QKD with multiple time-bin states encoded across each clock cycle, going beyond the conventional two-state encoding scheme. Good agreement is observed between the theoretical models considering realistic parameters and the experimental results with the eight-state encoded MDI-QKD achieving maximum SKR scaling with 3- and 2.63-times improvement across 2 and 50 km link length when compared to the conventional two-state qubit-based protocol. A comparison of the larger alphabet set implementation for MDI- and COW-QKD reveals that the SKR scales favorably for MDI-QKD with increasing encoding dimension, owing to the increased level-$d$ coincidence events allowing larger fraction of transmitted Z-basis states to contribute to useful key exchange. We also emphasize that the objective of the present work is aimed at demonstrating protocol-level advantage by expanding the alphabet set used for MDI-QKD and not optimized for achieving highest absolute SKR for MDI-QKD. It is pointed out that there are previous notable demonstrations of long-distance MDI-QKD achieving positive key exchange across 100s of km and SKR on the order of few kbps by employing superconducting nanowire single-photon detectors (SNSPDs), ultra-low loss fibers and highest clock rates [43]. In contrast, the present implementation is optimized for use of SPADs and standard single mode fibers. The SPAD-based implementations generally yield lower raw key rates compared to SNSPD-based ones. The proposed large alphabet encoding technique increasing the key rate is best suited for SPADs owing to their longer dead-times limiting the useful detection events. Consequently, this provides a practical solution for increasing secure key throughput in a cost-effective manner for short-to-medium range metropolitan links, where detector deadtime limits the raw key throughput and nicely complement the advances in long distance MDI-QKD solutions.

The SKR and corresponding error rates obtained in this study can be further improved by increasing the signal-to-noise ratio of the system, which in turn enhances the probability of generating high-fidelity level-$d$ coincidence events. Notably, the use of SNSPD can increase the overall detection rate while simultaneously reducing dark counts, after-pulsing, and detector deadtimes, thereby boosting the SKR, albeit at the cost of increased system complexity and cost. High-performance components such as, modulators with increased on-off extinction for precise state generation and low loss optical components, can further improve the key rates closer to the ideal theoretical limits, thereby achieving higher SKRs with larger alphabet sets. This study demonstrates the feasibility of enlarging the alphabet set for encoding in MDI-QKD offering a practical pathway for increasing secret-key rates compared to conventional two-state encoding scheme without requiring any hardware modification. The measurement device performs partial Bell-state projections onto two-dimensional states, thereby achieving measurement-device independence within the standard two-dimensional MDI-QKD security framework. The present implementation employs only two X-basis states for phase encoding, which was shown to be sufficient to demonstrate security under a prepare-and-measure framework [36] but does not constitute a complete high-dimensional MDI-QKD protocol involving all mutually unbiased basis states. Extending the scheme to incorporate the full X-basis set and developing the corresponding unconditional security proof remain important directions for future work. This work provides important insights into the scalability of MDI-QKD without requiring any hardware modifications thereby paving the way for realizing next-generation high-throughput QKD networks using existing QKD resources.

**Acknowledgements:** The authors acknowledge financial support from the Defense Research and Development Organization (DRDO), Government of India, under the Defense Industry-Academia Center of Excellence-funded project (No. DFTM/03/2023/P/40/JATC-P2QP-16/604/D(R&D)/2023) and Department of Science and Technology, Government of India, under the National Quantum Mission, Quantum Communication Technology-Hub.

## References

1. Rivest RL, Shamir A, Adleman L. A method for obtaining digital signatures and public-key cryptosystems. *Commun. ACM* 1978;21:120–6. https://doi.org/10.1145/359340.359342
2. National Institute of Standards and Technology. *Advanced Encryption Standard (AES)*. FIPS PUB 197, U.S. Department of Commerce; 2001. https://doi.org/10.6028/NIST.FIPS.197
3. Wolf R. Introduction Quantum Key Distribution. *Introduction* 2021;988:53–89.
4. Wright MA. The impact of quantum computing on cryptography. *Netw. Secur.* 2000;2000:13–5. https://doi.org/10.1016/S1353-4858(00)09027-9
5. Brassard G. Quantum computing: The end of classical cryptography? *ACM SIGACT News* 1994;25:15–21. https://doi.org/10.1145/190616.190617
6. Bennett CH, Brassard G, Mermin ND. Quantum cryptography without Bell's theorem. *Phys. Rev. Lett.* 1992;68:557–9. https://doi.org/10.1103/PhysRevLett.68.557
7. Shor PW. Algorithms for quantum computation: Discrete logarithms and factoring. In: *Proc. 35th Annu. Symp. Found. Comput. Sci.* IEEE; 1994. p. 124–34. DOI: 10.1109/SFCS.1994.365700
8. Hallgren S. Polynomial-time quantum algorithms for Pell's equation and the principal ideal problem. J. ACM 2007;54(1):1–19. https://doi.org/10.1145/1206035.1206039
9. Bennett CH, Brassard G. Quantum cryptography: Public key distribution and coin tossing. In: *Proc. IEEE Int. Conf. Comput. Syst. Signal Process.* 1984. p. 175–9.
10. Ekert AK. Quantum cryptography based on Bell's theorem. *Phys. Rev. Lett.* 1991;67:661–3. https://doi.org/10.1103/PhysRevLett.67.661
11. Lo HK, Chau HF. Unconditional security of quantum key distribution over arbitrarily long distances. *Science* 1999;283:2050–6. https://doi.org/10.1126/science.283.5410.2050
12. Shor PW, Preskill J. Simple proof of security of the BB84 quantum key distribution protocol. *Phys. Rev. Lett.* 2000;85:441–4. https://doi.org/10.1103/PhysRevLett.85.441
13. Gobby C, Yuan ZL, Shields AJ. Quantum key distribution over 122 km of standard telecom fiber. *Appl. Phys. Lett.* 2004;84:3762–4. https://doi.org/10.1063/1.1738173
14. Gisin N, Ribordy G, Tittel W, Zbinden H. Quantum cryptography. *Rev. Mod. Phys.* 2002;74:145–95. https://doi.org/10.1103/RevModPhys.74.145
15. Hadfield RH. Single-photon detectors for optical quantum information applications. *Nat. Photonics* 2009;3:696–705. https://doi.org/10.1038/nphoton.2009.230
16. Korzh B, et al. Provably secure and practical quantum key distribution over 307 km of optical fiber. *Nat. Photonics* 2015;9:163–8. https://doi.org/10.1038/nphoton.2014.327
17. Ursin R, et al. Entanglement-based quantum communication over 144 km. *Nat. Phys.* 2007;3:481–6. https://doi.org/10.1038/nphys629
18. Scarani V, et al. The security of practical quantum key distribution. *Rev. Mod. Phys.* 2009;81:1301–50. https://doi.org/10.1103/RevModPhys.81.1301
19. Makarov V. Controlling passively quenched single-photon detectors by bright light. *New J. Phys.* 2009;11:065003. https://doi.org/10.1088/1367-2630/11/6/065003
20. Qi B, Fung CHF, Lo HK, Ma X. Time-shift attack in practical quantum cryptosystems. *Quantum Inf. Comput.* 2007;7:73–82.
21. Jain N, et al. Trojan-horse attacks threaten the security of practical quantum cryptography. *New J. Phys.* 2011;13:123017. https://doi.org/10.1088/1367-2630/16/12/123030

22. Lydersen L, et al. Hacking commercial quantum cryptography systems by tailored bright illumination. *Nat. Photonics* 2010;4:686–9. https://doi.org/10.1038/nphoton.2010.214
23. Lo HK, Curty M, Qi B. Measurement-device-independent quantum key distribution. *Phys. Rev. Lett.* 2012;108:130503. https://doi.org/10.1103/PhysRevLett.108.130503
24. Rubenok A, Slater JA, Chan P, Lucio-Martinez I, Tittel W. Real-world two-photon interference and proof-of-principle quantum key distribution immune to detector attacks. *Phys. Rev. Lett.* 2013;111:130501. https://doi.org/10.1103/PhysRevLett.111.130501
25. Liu Y, et al. Experimental measurement-device-independent quantum key distribution. *Phys. Rev. Lett.* 2013;111:130502. https://doi.org/10.1103/PhysRevLett.111.130502
26. Tang Z, et al. Field test of measurement-device-independent quantum key distribution. *IEEE J. Sel. Top. Quantum Electron.* 2014;21:6600407. https://doi.org/10.1109/JSTQE.2014.2361796
27. Wang C, et al. Measurement-device-independent quantum key distribution over a 404 km optical fiber. *Phys. Rev. Lett.* 2017;118:200502. https://doi.org/10.1103/PhysRevLett.117.190501
28. Comandar LC, et al. Quantum key distribution without detector vulnerabilities using optically seeded lasers. Nat. Photonics 2016;10:312–315. https://doi.org/10.1038/nphoton.2016.50
29. Islam NT, et al. Provably secure and high-rate quantum key distribution with time-bin qudits. *Sci. Adv.* 2017;3:e1701491. https://doi.org/10.1126/sciadv.1701491
30. Sulimany K, Dudkiewicz R, Korenblit S, Eisenberg HS, Bromberg Y, Ben-Or M. Scrambled Time-Bin Encoding for Efficient High-Dimensional Quantum Key Distribution. In: Proc. CLEO: Applications & Technology (Optica Publishing Group) 2022, paper AM3D.2. https://doi.org/10.1364/CLEO_AT.2022.AM3D.2
31. Vagniluca I, et al. Efficient time-bin encoding for practical high-dimensional quantum key distribution. *Phys. Rev. Appl.* 2020;14:014051. https://doi.org/10.1103/PhysRevApplied.14.014051
32. Islam NT, Lim CCW, Cahall C, Kim J, Gauthier DJ. Scalable high-rate, high-dimensional time-bin encoding quantum key distribution. *Quantum Sci. Technol.* 2019;4:035008. https://doi.org/10.1088/2058-9565/ab21a4
33. Dellantonio L, Sørensen AS, Bacco D. High-dimensional measurement-device-independent quantum key distribution on two-dimensional subspaces. *Phys. Rev. A* 2018;98:062301. https://doi.org/10.1103/PhysRevA.98.062301
34. Brassard G, Lütkenhaus N, Mor T, Sanders BC. Limitations on practical quantum cryptography. *Phys. Rev. Lett.* 2000;85:1330. https://doi.org/10.1103/PhysRevLett.85.1330
35. Semenenko H, Sibson P, Hart A, Thompson MG, Rarity JG, Erven C. Chip-based measurement-device-independent quantum key distribution. *Optica* 2020;7:238–42. https://doi.org/10.1364/OPTICA.379679
36. Islam NT, Lim CCW, Cahall C, Kim J, Gauthier DJ. Securing quantum key distribution systems using fewer states. *Phys. Rev. A* 2018;97:042347. https://doi.org/10.1103/PhysRevA.97.042347
37. L. Vaidman, N. Yoran, Phys. Rev. A 1999, 59, 116. https://doi.org/10.1103/PhysRevA.59.116
38. J. Calsamiglia, N. Lütkenhaus, Appl. Phys. B 2001, 72, 67. https://doi.org/10.1007/s003400000484

39. Chan P, Slater JA, Lucio-Martinez I, Rubenok A, Tittel W. Modelling a measurement-device-independent quantum key distribution system. *Opt. Express* 2014;22:12716–36. https://doi.org/10.1364/OE.22.012716
40. Brádler K, Mirhosseini M, Fickler R, Broadbent A, Boyd RW. Finite-key security analysis for multilevel quantum key distribution. *New J. Phys.* 2016;18:073030. https://doi.org/10.1088/1367-2630/18/7/073030
41. Maity S, Prosad A, Natarajan H, Balaswamy V, Raghunathan V. Comparative Study of ASE-ASE and ASE-Laser Based Quantum Random Number Generators. IEEE Photonics J. 2024;16:7500110. https://doi.org/10.1109/JPHOT.2024.3373644
42. Gokul A, Natarajan H, Raghunathan V. Fiber-based higher dimensional quantum key distribution implementation using time-bin qudits. *Quantum Comput. Commun. Simul. IV, Proc. SPIE* 2024 ;12911 :1291105. https://doi.org/10.1117/12.3002155
43. Yin H-L et al. Measurement-device-independent quantum key distribution over a 404 km optical fiber. *Phys. Rev. Lett.* 2016;117:190501. https://doi.org/10.1103/PhysRevLett.117.190501
44. P. Zeng, H. Zhou, W. Wu, X. Ma, Mode-Pairing Quantum Key Distribution, Phys. Rev. X 12, 021041 (2022). https://doi.org/10.1038/s41467-022-31534-7